\documentclass[a4paper,12pt]{article}
\def\half{\frac{1}{2}}
\def\vf{\varphi}
\def\a{\alpha}
\def\b{\beta}
\def\p{\partial}
\def\w{\omega}

\title{On Pseudo-Hermitian Hamiltonians and Their Hermitian Counterparts}

\author{H.~F.~Jones\footnote{email:
{\tt h.f.jones@imperial.ac.uk }}  \\
{\small\it Physics Department, Imperial College,}\\
{\small\it London SW7 2BZ, United Kingdom.} }

\date{}

\begin{document}

\maketitle
\begin{abstract}
In the context of two particularly interesting non-Hermitian
models in quantum mechanics we
explore the relationship between the original Hamiltonian $H$
and its Hermitian counterpart $h$, obtained from $H$ by
a similarity transformation, as pointed out by Mostafazadeh.
In the first model, due to Swanson, $h$ turns out to be just a scaled
harmonic oscillator, which explains the form of its spectrum. However,
the transformation is not unique, which also means that the observables
of the original theory are not uniquely determined by $H$ alone.
The second model we consider is the original PT-invariant Hamiltonian, with potential
$V=igx^3$. In this case the corresponding $h$, which we are only able to construct
in perturbation theory, corresponds to a complicated velocity-dependent potential.
We again explore the relationship between the canonical variables $x$ and $p$
and the observables $X$ and $P$.
\end{abstract}
\section{Introduction}

There has recently been a great deal of interest in the properties
of non-Hermitian Hamiltonians, particularly those which possess
$PT$ symmetry, of which the prototype is the Hamiltonian
\begin{equation}\label{ix3}
H=\half(p^2+x^2)+igx^3,
\end{equation}
first studied in detail by Bender and Boettcher\cite{Bender1}, following an earlier
suggestion by Bessis.

This Hamiltonian was shown numerically to have a real, positive spectrum,
as indeed were its generalizations to
\begin{equation}\label{ixN}
H=\half(p^2+x^2)+g x^2(ix)^N.
\end{equation}
A rigorous proof of the reality of the spectrum was subsequently given by
Dorey et al.\cite{Dorey}.

In the intervening time many examples of non-Hermitian Hamiltonians were
found, often complex generalizations of well-known soluble potentials such
as the Morse potential, which all possessed real spectra for some range of
the parameters. However, the focus then moved on to more difficult problems
posed by such Hamiltonians, namely whether they possessed a consistent
interpretational framework. The problem arises because in such theories the natural metric in
the space of quantum mechanical states does not necessarily possess the attribute
of positive definiteness which is the basis of the probabilistic interpretation
of quantum mechanics.

In the context of $PT$-invariant theories a solution
was proposed by Bender et al.\cite{Bender2}, who introduced a new operator $C$
and a new scalar product, the $CPT$ scalar product, which was indeed positive definite.
This solves the problem in principle, but the difficulty is that the new product is
dynamically determined, that is, one needs to know the eigenvalues and eigenvectors
of the Hamiltonian in order to construct $C$. This can be done for soluble models, but
for the prototype Hamiltonian of Eq.~(\ref{ix3}) only a perturbative expansion for $C$
is available.

In a parallel development, Mostafazadeh\cite{Mostafazadeh} introduced the notion of
pseudo-Hermiticity. A Hamiltonian is said to be pseudo-Hermitian with respect to
a positive-definite, Hermitian operator $\eta$ if it satisfies
\begin{equation}
H^\dag=\eta H \eta^{-1}.
\end{equation}
In the case of $PT$-symmetric Hamiltonians the role of $\eta$ is played by
$PC$. In Ref.~\cite{Bender3} it was found convenient to write $C$ in the form
$C=e^Q P$, where $Q$ was a Hermitian operator satisfying $PQ=-QP$. Hence
in this case $\eta=e^{-Q}$, which is indeed a positive-definite Hermitian
operator.

The positive-definite metric takes the form
\begin{equation}\label{PDmetric}
\langle\langle \vf , \psi\rangle\rangle=\langle \vf, \eta\psi\rangle,
\end{equation}
where $\langle\phantom{xx}\rangle$ denotes the usual scalar product.

Further, the observables of the theory were identified as pseudo-Hermitian operators $A$
with respect to $\eta$. In the case of $PT$-symmetric theories where the
Hamiltonian is an even function of $p$, which includes the class of
Eq.~(\ref{ixN}), this coincides with the definition\cite{Bender2p} that
$A$ must satisfy $\tilde{A}=(CPT)A(CPT)$.

Mostafazadeh went on to show that under a similarity transformation
implemented by $\rho=\sqrt{\eta}$ such a Hamiltonian is equivalent
to a Hermitian Hamiltonian $h$, according to
\begin{equation}
H=\rho^{-1} h \rho.
\end{equation}
Again, a similar relation holds for observables in general:
if $a$ is an observable in the Hermitian theory described by $h$,
the corresponding observable in the pseudo-Hermitian theory is
\begin{equation}
A=\rho^{-1}a\rho.
\end{equation}

In the present paper we wish to explore these relationships in
detail in two models. One, initially presented by
Swanson\cite{Swanson}, is a soluble model which can be transformed
by a similarity transformation (in fact a whole class of such
transformations) to a simple harmonic oscillator. Here we discuss
the different possible similarity transformations, and in the
simplest case, where $\eta=\eta(x)$, identify the observables. The
second model, which can only be treated in perturbation theory, is
the original $igx^3$ Hamiltonian of Eq.~(\ref{ix3}). In this case
we construct $h$ to order $g^4$ and the
observables to order $g^2$. The resulting $h$ is
a complicated, momentum-dependent object, in contrast to the simple form of
$H$. This means that although the two theories are formally equivalent,
the non-Hermitian $H$ is the only practical starting point.


\section{The Swanson Hamiltonian}

An interesting Hamiltonian, which is $PT$-symmetric, but not
symmetric, is that considered by Swanson\cite{Swanson}:
\begin{equation}\label{Swanson}
H=\w a^\dag a +\a a^2 + \b {\a^\dag}^2,
\end{equation}
where $a$ and $a^\dag$ are harmonic oscillator annihilation and
creation operators for unit frequency and $\w$, $\a$ and $\b$ are real constants.
This Hamiltonian has a real, positive spectrum in a certain range of
the parameters.

In fact for $\omega>\a+\b$, the spectrum of Eq.~(\ref{Swanson}) is
that of the simple harmonic oscillator with frequency
$\Omega=\sqrt{\w^2-4\a\b}$. Swanson showed this by constructing
a transformation operator $U (=\eta)$ of Bogoliubov type which
reduced the original problem to that of the
simple harmonic oscillator. This gave the following form\footnote{taking
the $g_i$ as real} for $U$:
\begin{equation}\label{SwansonU}
 U=\exp\left\{\half\left(\frac{g_3}{g_1}-\frac{g_2}{g_4}\right){a^\dag}^2\right\}
 \exp\left(\half w d^2\right)\exp(cd\ln z),
\end{equation}
where $w=(g_3g_4-g_1g_2)/g_4^2$,$\;$ $z=g_4/g_1$, and $c$ and
$d$ are Bogoliubov transforms of $a$ and $a^\dag$:
\begin{eqnarray*}
c&=&g_1a^\dag-g_3a,\cr d&=&g_4a-g_2a^\dag.
\end{eqnarray*}
The $g_i$ are
subject to the three conditions
\begin{eqnarray*} g_1g_4-g_2g_3&=&1,\cr
g_2g_4\w+g_2^2\a+g_4^2\b&=&0,\cr g_1g_3\w+g_1^2\a+g_3^2\b&=&0,
\end{eqnarray*}
which means that there is a one-parameter family of solutions,
depending on $g_1$, say. Geyer et al.\cite{Geyer} noted this
non-uniqueness of $U$, in contrast to the uniquely defined
operator $C$, or $Q$, of refs.~\cite{Bender2, Bender3}, and
proposed that the ambiguity could be removed by the requirement
that not only the Hamiltonian but a given observable (or in
general an ``irreducible set of observables") should be
pseudo-Hermitian with respect to $\eta$.

In fact what this amounts to in this case is that $\eta$ is a function of that
particular observable. A very simple form of $\eta$ can be found\cite{Geyer}
by requiring it to be a function of the number operator $N=a^\dag a$.
In fact, with $S (=\rho=\eta^\half)$ given by
\begin{equation}\label{Geyer}
S=\exp\left[\frac{1}{4}N\ln(\a/\b)\right],
\end{equation}
it is easy to see, using the commutation relations
 $[N,A]=2B$, $[N,B]=2A$, where $A:={a^\dag}^2+a^2$,
$B:={a^\dag}^2+a^2$, that
\begin{equation}
h=SHS^{-1}=\half p^2(\w-2\sqrt{\a\b})+\half x^2(\w+2\sqrt{\a\b}),
\end{equation}
a scaled harmonic oscillator with frequency $\Omega$.

The condition $[S,N]=0$ gives the additional constraint $g_1g_3=g_2g_4$
on the parameters $g_i$: however, it is still not easy to see the
equivalence between the three-exponential form of Swanson, Eq.~(\ref{SwansonU}) and the
single-exponential form of Eq.~(\ref{Geyer}).

While this transformation is adequate to obtain the spectrum
of $H$, it is not suitable for calculations in wave mechanics,
where explicit eigenfunctions are needed. An alternative transformation,
which immediately gives the form of the wave functions, is obtained by
choosing $\eta$ to be a function of $x$. Indeed, it is easily seen that
the required form of $\rho$ is
\begin{equation}\label{HFJ}
\rho=\exp\left[{\half \lambda x^2}\right],
\end{equation}
where
\begin{eqnarray*}
\lambda=\frac{\b-\a}{\w-\a-\b}.
\end{eqnarray*}
By virtue of the commutation relations $[x^2,A]=2B$, $[x^2,B]=2A+4C$, $[x^2,C]=-B$,
where $C:=N+\half$, the similarity transformation $\rho H \rho^{-1}$ now gives
\begin{equation}\label{HFJh}
h=\rho H \rho^{-1}=\half p^2(\w-\a-\b)+\half x^2\frac{\w^2-4\a\b}{\w-\a-\b},
\end{equation}
a different scaled harmonic oscillator with the same frequency $\Omega$.

This transformation corresponds to the method of reducing the original Schr\"odinger
differential equation for $\psi$:
\begin{eqnarray*}
\left[\half\w\left(x^2-\frac{d^2}{dx^2}\right)+\half(\a+\b)\left(x^2+\frac{d^2}{dx^2}\right)
+(\a-\b)x\frac{d}{dx}\right]\psi=E\psi,
\end{eqnarray*}
to that of a simple harmonic oscillator for $\vf$ by writing $\psi=W\vf$
and choosing $W$ so that there are no linear derivatives acting on $\vf$.
The resulting condition on $W$ is $(\w-\a-\b)W'+(\b-\a)xW=0$, which gives
$W=\rho^{-1}$. The resulting wave functions are
\begin{equation}
\psi_n={\cal N}_n e^{-\half x^2(\lambda+\mu^2)}H_n(\mu x),
\end{equation}
where
\begin{eqnarray*}
\mu=\frac{(\w^2-4\a\b)^{\frac{1}{4}}}{(\w-\a-\b)^\half},
\end{eqnarray*}
the $H_n$ are the Hermite polynomials and ${\cal N}_n$ is the
appropriate normalization factor. Clearly these are not orthonormal as
they stand; rather they are orthonormal with respect to the weight factor
$\eta=\rho^2=e^{\lambda x^2}$. That is,
\begin{equation}
\int \psi^*_m(x)e^{\lambda x^2}\psi_n(x)\ dx = \delta_{mn}\ ,
\end{equation}
in accordance with Eq.~(\ref{PDmetric}).

If one takes the point of view that the original Hamiltonian $H$ is obtained
by the inverse similarity transformation from the $h$ of Eq.~(\ref{HFJh}),
then the observables of the non-Hermitian $H$ theory are obtained by the same
inverse transformation on those of $h$, which in addition to $h$ are $x$ and $p$.
Thus
\begin{eqnarray}
X:=e^{-\half\lambda x^2}x\ e^{\half\lambda x^2}&=& x ,\cr
P:=e^{-\half\lambda x^2}p\ e^{\half\lambda x^2}&=& p-i\lambda x .
\end{eqnarray}
Equally $H$ can be written in the form of Eq.~(\ref{HFJh}), with $p$ replaced
by $P$. This approach, namely deriving a non-Hermitian Hamiltonian by the
above transformation of $p$, was in fact originally taken by
by Ahmed\cite{Ahmed} before the paper of Swanson.

However, a rather puzzling situation arises, in that what we define as the observables
associated with $H$ depends on the particular transformation $\rho$ that is used to
convert it to a simple harmonic oscillator. Thus, apart from the transformation used
by Geyer et al. and that just discussed, it would be equally possible to take $\rho$ as
a function of $p$ alone. In that case we would have simple wave-functions in momentum space,
and the observables would be $p$ and a transformed version of $x$. As already discussed above,
there is in fact a one-parameter family of transformations, and hence of observables.

\section{The $igx^3$ Theory}

For the Hamiltonian of Eq.~(\ref{ix3}) the $Q$ operator has been constructed\cite{Bender2}
up to $O(g^7)$ in the form $Q=\sum_r g^rQ_r$. To first order\footnote{In fact
this result is accurate up to second order: $Q$ contains only odd
powers of $g$.}
\begin{equation}\label{Q1}
Q_1=-\frac{4}{3}p^3-2xpx.
\end{equation}
We are thus in a position to construct $h$, which to this order is given by
\begin{equation}
h=e^{-\half gQ_1}H e^{\half gQ_1}.
\end{equation}
By virtue of the property
$
[Q_1,H_0]=2H_1,
$
where $H_0\equiv \half(p^2+x^2)$
and $H_1\equiv ix^3$, this becomes
\begin{eqnarray}
h(x,p)&=&H_0-\frac{1}{4}g^2[Q_1,H_1]\cr&&\cr
&=&H_0+3g^2\left(\half x^4+S_{2,2} -\frac{1}{6}\right)+O(g^4),
\end{eqnarray}
where $S_{2,2}=(x^2p^2+xp^2x+p^2x^2)/3$. As indicated, the next
correction is of order $g^4$ by virtue of the structure of the
commutation relations of the $Q_r$.

This result is interesting in several respects. Firstly it is already
quite complicated, compared with the simple form of $H$, and that complication
only increases in higher orders. Secondly it has an $x^4$ component with
a positive sign, as for a conventional quartic oscillator. Thirdly it is momentum-dependent,
containing a number of terms involving $p$.

The calculation can be continued, using
\begin{eqnarray*}
Q_3={ \frac{128}{15}}p^5+\frac{40}{3}S_{3,2}+8S_{1,4}-12p,
\end{eqnarray*}
where the $S_{m,n}$ are fully symmetrized polynomials of degree $m$ in $x$
and $n$ in $p$. The fourth-order contribution is
\begin{eqnarray*}
h_4=g^4[-(7/2)x^6-(51/2)S_{2,4}-
36S_{4,2}+2p^6+(15/2)x^2+27p^2],
\end{eqnarray*}
which now has a negative coefficient for the $x^6$ term,
but also contains a term in $p^6$. If we were able to sum
up the perturbation series, the higher powers of $p=-i\p/\p x$ would
ultimately produce a non-local function.
Clearly this is not a Hamiltonian that one would
have contemplated in its own regard, were it not derived from Eq.~(\ref{ix3}).
It is for this reason that we disagree with the contention of Mostafazadeh\cite{Mostafazadeh1} that
``a consistent probabilistic PT-symmetric quantum theory is doomed to reduce to
ordinary QM".

Turning to the question of the observables of the system, these are obtainable
from those of the Hermitian theory, namely $x$ and $p$, by the transformations
\begin{eqnarray}
X&=&e^{\half Q}x\ e^{-\half Q} \cr
P&=&e^{\half Q}p\ e^{-\half Q}.
\end{eqnarray}
To second order these are
\begin{eqnarray}
X&=&x+ig(x^2+2p^2) +g^2(-x^3+2pxp),\cr
P&=&p-ig(xp+px)+g^2(2p^3-xpx).
\end{eqnarray}
Again, these calculations can be carried out to higher order, but the results
are not particularly illuminating.

It is, however, interesting to compare the ground-state expectation values
\begin{eqnarray*}
\langle\langle\psi_0, X\psi_0\rangle\rangle= \langle \psi_0, e^{-\half Q} x e^{-\half Q}\psi_0\rangle=0
\end{eqnarray*}
and
\begin{eqnarray*}
\langle\langle\psi_0, x\psi_0\rangle\rangle= \langle \psi_0, e^{-Q} x \psi_0\rangle=-\frac{3}{2}ig + O(g^3)\ .
\end{eqnarray*}
The first must be real, and is in fact zero by symmetry, whereas the second is pure imaginary. This is
unacceptable for an observable in quantum mechanics. In the generalization to quantum field
theory, however, where $x(t)\to \vf(\mathbf{x},t)$, the field itself is not necessarily an
observable, so a non-vanishing expectation value may be acceptable.

Note that $Q$ itself is an observable, since it is Hermitian and commutes with itself.
It also has the property that
\begin{equation}
Q(x,p)=e^{\half Q}Q(x,p)\,e^{-\half Q}= Q(X,P)\, .
\end{equation}
That is, $Q$, originally written as a function of $x$ and $p$, is in fact
the same function of the observables $X$ and $P$.

Since $X$ and $P$ are the observables, it might be tempting to express $H$
in terms of them, instead of the original $x$ and $p$. Unfortunately this
does not lead to any appreciable simplification,  because in fact
\begin{eqnarray*}
H=e^{\half Q}h(x,p)e^{-\half Q}= h(X,P).
\end{eqnarray*}
That is, the initial, non-Hermitian Hamiltonian $H$, when expressed in terms
of the observables $X$ and $P$, is of the same form as $h$: a complicated, momentum-dependent
function.


\section{Discussion}

In the context of two particularly interesting models, we have discussed the relation between the
original non-Hermitian Hamiltonian $H$ and its Hermitian counterpart $h$, and have exhibited
the observables of the theory. In the case of the Swanson Hamiltonian of
Eq.~(\ref{Geyer}), there is a one-parameter choice for the transformation operator $\eta$,
and correspondingly the observables of the theory are not determined uniquely by the Hamiltonian
but depend on that choice. For the $ix^3$ model of Eq.~(\ref{ix3}) we explicitly constructed the
corresponding Hermitian Hamiltonian in perturbation theory, noting its complicated, momentum-dependent character.
We constructed the observables $X$ and $P$ in perturbation theory and discussed their relation to
the canonical $x$, $p$.

\section*{Acknowledgements}

I am grateful to the organizers of the 2nd International Workshop on
Pseudo-Hermitian Hamiltonians in Quantum Physics (Prague, 2004) for providing a stimulating
environment, to Prof.~H.~Geyer for bringing my attention to the Swanson Hamiltonian
and its associated problems, and to Prof.~C.~M.~Bender for extremely helpful discussions.



\end{document}